\documentclass[12pt]{article}%
\usepackage{amsmath}
\usepackage{graphicx}%
\usepackage{amsfonts}%
\usepackage{amssymb}
\newtheorem{theorem}{Theorem}
\newtheorem{acknowledgement}[theorem]{Acknowledgement}

\newtheorem{example}[theorem]{Example}

\newtheorem{proposition}[theorem]{Proposition}
\newtheorem{remark}[theorem]{Remark}

\newenvironment{proof}[1][Proof]{\textbf{#1.} }{\ \rule{0.5em}{0.5em}}
\begin{document}

\title{Quaternionic reformulation of Maxwell's equations for inhomogeneous media and
new solutions}
\author{Vladislav V. Kravchenko\\Departamento de Telecomunicaciones,\\Escuela Superior de Ingenier\'{\i}a Mec\'{a}nica y El\'{e}ctrica,\\Instituto Polit\'{e}cnico Nacional, C.P. 07738, D.F., MEXICO\\e-mail: vkravche@maya.esimez.ipn.mx}
\maketitle

\begin{abstract}
We propose a simple quaternionic reformulation of Maxwell's equations for
inhomogeneous media and use it in order to obtain new solutions in a static case.

\end{abstract}



\section{Introduction}

The algebra of quaternions was applied to the study of Maxwell's equations
starting from the work of J. C. Maxwell himself. The standard reference for
the quaternionic reformulation of Maxwell's equations is the work
\cite{Imaeda}, where the Maxwell equations for vacuum were written in a simple
and compact form. In this relation we should mention also the earlier article
\cite{Shn}. Some new integral representations for electromagnetic quantities
based on the idea of quaternionic diagonalization of Maxwell's equations for
homogeneous media were obtained in \cite{Krdep} (see also \cite[Chapter 2]%
{KS}). A review of different applications of quaternionic analysis to the
Maxwell equations can be found in \cite{GS2}. In the recent work
\cite{KrZAAel} with the aid of quaternionic analysis techniques the Maxwell
equations for inhomogeneous but slowly changing media were diagonalized and
new solutions obtained. Nevertheless even the question: how to write the
Maxwell equations for arbitrary inhomogeneous media in a compact quaternionic
form remained open. In the present work we propose such a reformulation and
use it for obtaining new results in the static case.

\bigskip

\section{Preliminaries}

We denote by ${\mathbb{H}}({\mathbb{C}})$ the algebra of complex quaternions
(=biquaternions). The elements of ${\mathbb{H}}({\mathbb{C}})$ are represented
in the form $q=\sum_{k=0}^{3}q_{k}i_{k}$, where $q_{k}\in\mathbb{C}$, $i_{0}$
is the unit and $i_{k}$, $k=1,2,3$ are standard quaternionic imaginary units.
We will use also the vector representation of complex quaternions:
$q=q_{0}+\overrightarrow{q}$, where $\overrightarrow{q}=\sum_{k=1}^{3}%
q_{k}i_{k}$. The vector parts of complex quaternions we identify with vectors
from ${\mathbb{C}}^{3}$. The product of two biquaternions can be written in
the following form%

\[
p\cdot q=p_{0}q_{0}-<\overrightarrow{p},\overrightarrow{q}>+\left[
\overrightarrow{p}\times\overrightarrow{q}\right]  +p_{0}\overrightarrow
{q}+q_{0}\overrightarrow{p},
\]
where $<\overrightarrow{p},\overrightarrow{q}>$ and $\left[  \overrightarrow
{p}\times\overrightarrow{q}\right]  $ denote the usual scalar and vector
products respectively.

We will use the following notations for the operators of multiplication from
the left-hand side and from the right-hand side%
\[
^{p}Mq=p\cdot q\quad\text{and\quad}M^{p}q=q\cdot p.
\]

\begin{remark}
\label{scalar}The scalar product of the vectors $\overrightarrow{p}$ and
$\overrightarrow{q}$ can be represented as follows%
\[
<\overrightarrow{p},\overrightarrow{q}>=-\frac{1}{2}(^{\overrightarrow{p}%
}M+M^{\overrightarrow{p}})\overrightarrow{q}.
\]
\end{remark}

On the set of differentiable ${\mathbb{H}}({\mathbb{C}})$-valued functions the
Moisil-Theodoresco operator is defined by the expression $Df=\sum_{k=1}%
^{3}i_{k}\partial_{k}f$. In a vector form this expression can be written as
follows%
\[
Df=-\operatorname*{div}\overrightarrow{f}+\operatorname*{grad}f_{0}%
+\operatorname*{rot}\overrightarrow{f},
\]
where the first term is the scalar part of the biquaternion $Df$ and the last
two terms represent its vector part.

Let us note some simple properties of the operator $D$ which will be used in
this work. Let $\varphi$ be a scalar function and $f$ be an ${\mathbb{H}%
}({\mathbb{C}})$-valued function. Then
\[
D(\varphi\cdot f)=D\varphi\cdot f+\varphi\cdot Df,
\]
and%
\begin{equation}
(D-\frac{\operatorname{grad}\varphi}{\varphi})f=\varphi D(\varphi^{-1}f).
\label{D-grad}%
\end{equation}

\bigskip

\section{Quaternionic reformulation of Maxwell's equations}

The Maxwell equations for complex amplitudes of the time-harmonic
electromagnetic field have the form%
\begin{equation}
\operatorname*{div}(\varepsilon\overrightarrow{E})=\rho,\qquad
\operatorname*{div}(\mu\overrightarrow{H})=0, \label{M1}%
\end{equation}%
\begin{equation}
\operatorname*{rot}\overrightarrow{H}=i\omega\varepsilon\overrightarrow
{E}+\overrightarrow{j},\qquad\operatorname*{rot}\overrightarrow{E}=-i\omega
\mu\overrightarrow{H}, \label{M2}%
\end{equation}
where $\overrightarrow{E}$ and $\overrightarrow{H}$ are ${\mathbb{C}}^{3}%
$-valued functions; $\omega$ is the frequency; $\varepsilon$ is the
permittivity and $\mu$ is the permeability of the medium. We suppose
$\varepsilon$ and $\mu$ to be two times differentiable complex valued
functions with respect to each coordinate $x_{k}$, $k=1,2,3$. Note that they
are always different from zero.

The equations (\ref{M1}) can be rewritten as follows%
\[
\operatorname*{div}\overrightarrow{E}+<\frac{\operatorname{grad}\varepsilon
}{\varepsilon},\overrightarrow{E}>=\frac{\rho}{\varepsilon}%
\]
and%
\[
\operatorname*{div}\overrightarrow{H}+<\frac{\operatorname{grad}\mu}{\mu
},\overrightarrow{H}>=0.
\]
Combining these equations with (\ref{M2}) we have the Maxwell equations in the
form%
\[
D\overrightarrow{E}=<\frac{\operatorname{grad}\varepsilon}{\varepsilon
},\overrightarrow{E}>-i\omega\mu\overrightarrow{H}-\frac{\rho}{\varepsilon}%
\]
and%
\[
D\overrightarrow{H}=<\frac{\operatorname{grad}\mu}{\mu},\overrightarrow
{H}>+i\omega\varepsilon\overrightarrow{E}+\overrightarrow{j}.
\]
Taking into account Remark \ref{scalar} we rewrite them as follows%
\[
(D+\frac{1}{2}\frac{\operatorname{grad}\varepsilon}{\varepsilon}%
)\overrightarrow{E}=-\frac{1}{2}M^{\frac{\operatorname{grad}\varepsilon
}{\varepsilon}}\overrightarrow{E}-i\omega\mu\overrightarrow{H}-\frac{\rho
}{\varepsilon}%
\]
and%
\[
(D+\frac{1}{2}\frac{\operatorname{grad}\mu}{\mu})\overrightarrow{H}%
=-\frac{1}{2}M^{\frac{\operatorname{grad}\mu}{\mu}}\overrightarrow{H}%
+i\omega\varepsilon\overrightarrow{E}+\overrightarrow{j}.
\]
Due to (\ref{D-grad}) we obtain%
\[
\frac{1}{\sqrt{\varepsilon}}D(\sqrt{\varepsilon}\cdot\overrightarrow
{E})+\overrightarrow{E}\cdot\overrightarrow{\varepsilon}=-i\omega
\mu\overrightarrow{H}-\frac{\rho}{\varepsilon}%
\]
and%
\[
\frac{1}{\sqrt{\mu}}D(\sqrt{\mu}\cdot\overrightarrow{H})+\overrightarrow
{H}\cdot\overrightarrow{\mu}=i\omega\varepsilon\overrightarrow{E}%
+\overrightarrow{j},
\]
where
\[
\overrightarrow{\varepsilon}=\frac{\operatorname{grad}\sqrt{\varepsilon}%
}{\sqrt{\varepsilon}}\qquad\text{and\qquad}\overrightarrow{\mu}%
=\frac{\operatorname{grad}\sqrt{\mu}}{\sqrt{\mu}}.
\]
Introducing the notations%
\[
\overrightarrow{\mathcal{E}}=\sqrt{\varepsilon}\overrightarrow{E}%
,\qquad\overrightarrow{\mathcal{H}}=\sqrt{\mu}\overrightarrow{H}%
\quad\text{and\quad}k=\omega\sqrt{\varepsilon\mu},
\]
we finally arrive at the following system%
\begin{equation}
(D+M^{\overrightarrow{\varepsilon}})\overrightarrow{\mathcal{E}}%
=-ik\overrightarrow{\mathcal{H}}-\frac{\rho}{\sqrt{\varepsilon}} \label{Mq1}%
\end{equation}
and%
\begin{equation}
(D+M^{\overrightarrow{\mu}})\overrightarrow{\mathcal{H}}=ik\overrightarrow
{\mathcal{E}}+\sqrt{\mu}\overrightarrow{j}, \label{Mq2}%
\end{equation}
which is equivalent to (\ref{M1}), (\ref{M2}). This pair of equations is the
quaternionic reformulation of the Maxwell equations for inhomogeneous media.

The operator $D+M^{\alpha}$, with $\alpha$ being a constant complex quaternion
was studied in detail in \cite{KS}. Note that $\overrightarrow{\varepsilon}$
and $\overrightarrow{\mu}$ are constants if $\varepsilon$ and $\mu$ are
functions of the form $\exp(ax_{1}+bx_{2}+cx_{3}+d)$ with constant $a$, $b$
and $c$. For the case when $\alpha$ is not a constant there were proposed some
classes of exact solutions in \cite{KrZAADir}, \cite{KrChemnitz} and
\cite{KR}. In the same articles the reader can see that the classical Dirac
operator with different potentials is closely related to the operator
$D+M^{\alpha}$. A simple matrix transform proposed in \cite{Krbag} turns the
classical Dirac operator into the operator $D+M^{\alpha}$, where $\alpha$
contains the mass and the energy of the particle as well as the terms
corresponding to potentials.

In a static case ($\omega=0$) we arrive at the equations%
\[
(D+M^{\overrightarrow{\varepsilon}})\overrightarrow{\mathcal{E}}=-\frac{\rho
}{\sqrt{\varepsilon}},
\]
and%
\[
(D+M^{\overrightarrow{\mu}})\overrightarrow{\mathcal{H}}=\sqrt{\mu
}\overrightarrow{j}.
\]
Thus we are interested in the solutions for the operator $D+M^{\overrightarrow
{\alpha}}$, where the complex quaternion $\overrightarrow{\alpha}$ has the
form $\overrightarrow{\alpha}=(\operatorname{grad}\varphi)/\varphi$ and the
function $\varphi$ is different from zero. Note that due to (\ref{D-grad}) the
operator $D+^{\overrightarrow{\alpha}}M$ permits a complete study which can be
found in \cite{Sproessig}, because it practically reduces to the operator $D$.
In the case of the operator $D+M^{\overrightarrow{\alpha}}$ the situation as
we will see later on is quite different.

Consider the equation%
\begin{equation}
(D+M^{\overrightarrow{\alpha}})\overrightarrow{f}=0. \label{mainhom}%
\end{equation}
Denote
\[
v=\frac{\Delta\varphi}{\varphi}.
\]
In other words $\varphi$ is a solution of the Schr\"{o}dinger equation%
\begin{equation}
-\Delta\varphi+v\varphi=0. \label{Schrodinger}%
\end{equation}

\begin{proposition}
\label{Prop}Let $\psi$ be another solution of (\ref{Schrodinger}). Then the
function
\begin{equation}
\overrightarrow{f}=(D-\overrightarrow{\alpha})\psi\label{Darboux}%
\end{equation}
is a solution of (\ref{mainhom}).
\end{proposition}

\begin{proof}
The proof consists of a simple calculation. Consider%
\begin{align*}
D\overrightarrow{f}  &  =-\Delta\psi-D\psi\cdot\frac{D\varphi}{\varphi}%
-\psi\cdot D(\frac{D\varphi}{\varphi})\\
&  =-v\psi-D\psi\cdot\frac{D\varphi}{\varphi}+\psi\cdot\left(  \frac{D\varphi
}{\varphi}\right)  ^{2}+\psi\frac{\Delta\varphi}{\varphi}\\
&  =-(D\psi-\frac{D\varphi}{\varphi}\cdot\psi)\frac{D\varphi}{\varphi
}=-\overrightarrow{f}\cdot\overrightarrow{\alpha}.
\end{align*}
\end{proof}

This proposition gives us the possibility to reduce the solution of
(\ref{mainhom}) to the Schr\"{o}dinger equation (\ref{Schrodinger}). Moreover,
if $\psi$ is a fundamental solution of the Schr\"{o}dinger operator:%
\[
(-\Delta+v)\psi=\delta,
\]
then the function $\overrightarrow{f}$ defined by (\ref{Darboux}) is a
fundamental solution of the operator $D+M^{\overrightarrow{\alpha}}$ that can
be seen following the proof of Proposition \ref{Prop}.

\begin{remark}
The proposition \ref{Prop} is closely related to the factorization of the
Schr\"{o}dinger operator proposed in \cite{Swansolo}, \cite{Swan}. Namely, for
a scalar function $u$ we have the equality%
\[
(D+M^{\alpha})(D-M^{\alpha})u=(-\Delta+v)u,
\]
if the complex quaternionic function $\alpha$ satisfies the equation%
\begin{equation}
D\alpha+\alpha^{2}=-v. \label{Riccati}%
\end{equation}
It is easy to check that for $\alpha=(\operatorname{grad}\varphi)/\varphi$ the
equation (\ref{Riccati}) is equivalent to (\ref{Schrodinger}). Equation
(\ref{Riccati}) can be considered as a natural generalization of the ordinary
differential Riccati equation. In \cite{KKW} and \cite{W} the corresponding
generalizations of the well known Euler theorems for the Riccati equation were obtained.
\end{remark}

Let us consider the following simple example of application of the Proposition
\ref{Prop}.

\begin{example}
Consider equation (\ref{mainhom}) in some domain $\Omega\subset\mathbb{R}^{3}$
and let $\Delta\varphi/\varphi=-c^{2}$ in $\Omega$, where $c$ is a complex
constant. In this case we are able to construct a fundamental solution for the
operator $D+M^{\overrightarrow{\alpha}}$. Denote%
\[
\psi(x)=\frac{e^{ic\left|  x\right|  }}{4\pi\left|  x\right|  }.
\]
This is a fundamental solution of the operator $-\Delta-c^{2}$. Then the
fundamental solution of $D+M^{\overrightarrow{\alpha}}$ is constructed as
follows%
\[
\overrightarrow{f}(x)=(D-\frac{\operatorname{grad}\varphi(x)}{\varphi
(x)})\frac{e^{ic\left|  x\right|  }}{4\pi\left|  x\right|  }=(-\frac{x}%
{\left|  x\right|  ^{2}}+ic\frac{x}{\left|  x\right|  }%
-\frac{\operatorname{grad}\varphi(x)}{\varphi(x)})\cdot\frac{e^{ic\left|
x\right|  }}{4\pi\left|  x\right|  },
\]
where $x=\sum_{k=1}^{3}x_{k}i_{k}$.
\end{example}

Note that it is not clear how to obtain this result for the Maxwell operators
$D+M^{\overrightarrow{\varepsilon}}$ and $D+M^{\overrightarrow{\mu}}$ using
other known methods.

\begin{acknowledgement}
The author wishes to express his gratitude to Prof. Vladimir Rabinovich for
useful discussions.

This work was supported by CONACYT, project 32424-E.
\end{acknowledgement}

\end{document}